\documentclass{article}

\usepackage{amsfonts,amsmath}

\newtheorem{lemma}{Lemma}
\newtheorem{theorem}{Theorem}

\def\R{{\mathbb R}}

\begin{document}

\title{Two-dimensional {S}chr{\"o}dinger operators with fast decaying
              potential and multidimensional {$L_2$}-kernel\thanks{
The authors acknowledges partial financial support from RFBR (grants
06-01-00094a, 06-01-00814a), complex integration project 2.15 SB RAS
Max Plank Mathematical Institute (I.A.T.) and DFG Research Unit 565
``Polyhedral Surfaces'' (TU-Berlin, S.P.Ts.)}}
\author{I.A.Taimanov \thanks{
Sobolev Institute of Mathematics, 630090 Novosibirsk;
taimanov@math.nsc.ru},\ \ S.P.Tsarev
\thanks{Institut f\"ur Mathematik,
 Technische Universit\"at Berlin, Germany,
\& Krasnoyarsk State Pedagogical University, Lebedevoi,, 89, 660049,
Krasnoyarsk; sptsarev@mail.ru}}
\date{}
\maketitle

In this note using Moutard transformations we show how explicit
examples of two-dimensional Schr\"odinger operators $L = - \Delta
+ u(x,y)$ with fast decaying potential and multidimensional
{$L_2$}-kernel may be constructed. In the explicit examples below
the potential $u(x,y)$ and square-summable solutions $\psi(x,y)$
of the equation $L\psi = 0$ are smooth rational functions of $x$
and $y$. In the best case constructed below, the potential $u$ and
the eigenfunctions $\psi$ decay as $1/r^8$ and $1/r^3$
respectively, where $r^2 = x^2 + y^2$. The potentials constructed
are exactly solvable at the zero energy level, in the sense that
all solutions of $L\psi=0$ can be built from harmonic functions
and their construction requires only quadratures.

For operators with  so fast decaying potentials a suitable
spectral theory is known  (\cite{F,NK}). In dimension one
existence of square-summable eigenfunctions at zero energy level
is impossible. The inverse spectral problem for multidimensional
Schr\"odinger operators at a fixed energy level was considered in
\cite{DKN} for the first time, where the case of double periodic
potential was investigated. For fast decaying potentials this
problem was studied mostly at positive energy levels \cite{GM}, or
below the ground state \cite{GN}.

The Moutard transformation for the two-dimensional Schr\"odinger
operator
$$
L = - \Delta+ u = -\left(\frac{\partial^2}{\partial
x^2}+\frac{\partial^2}{\partial y^2}\right) + u(x,y)
$$
produces another Schr\"odinger operator $\widetilde{L}$ with the
potential $\widetilde{u}$:
$$
\widetilde{L} = -\Delta + \widetilde{u} = -\Delta + (u - 2\Delta
\log \omega),
$$
where $\omega$ is a solution of the equation $L \omega = 0$. From
this we have $\widetilde{u} = 2 \frac{\omega_x^2 +
\omega_y^2}{\omega^2} - u$. For every solution  $\varphi$ of the
equation $L\varphi=0$ one can  construct a solution
$\widetilde{\varphi}$ of the equation $\widetilde{L}
\widetilde{\varphi}=0$:
$$
(\omega \widetilde{\varphi})_x = - \omega^2
\left(\frac{\varphi}{\omega}\right)_y, \ \ \
(\omega\widetilde{\varphi})_y =
\omega^2\left(\frac{\varphi}{\omega}\right)_x.
$$
The construction of $\widetilde{\varphi}$ from $\varphi$ is
reduced to quadratures and the constructed function
$\widetilde{\varphi}$ is uniquely determined up to terms of the form
$C\omega^{-1}$, where $C = \mathrm{const}$. Keeping in mind this
ambiguity we will denote hereafter the result of the Moutard
transformation as $M_\omega$: $ \widetilde{u} = M_\omega(u)$,
$\widetilde{\varphi} = M_\omega(\varphi)$. Note that $\widetilde{L}
\omega^{-1} = 0$.

For the case of a one-dimensional potential $u=u(x)$ and
$\omega(x,y) = f(x)e^{ky}$ one has $L = -\frac{\partial ^2}{\partial
y^2} + A^\top A + k^2$, where $A = -\frac{\partial}{\partial x} +
\frac{f_x}{f}, A^\top = \frac{\partial}{\partial x} +
\frac{f_x}{f}$, so the Moutard transformation will be reduced to the Darboux
transformation of the one-dimensional operator $L^\prime =
-\frac{\partial^2}{\partial x^2} + (u+k^2)$, namely $L
\longrightarrow \widetilde{L} = -\frac{\partial^2}{\partial y^2} + A
A^\top + k^2.$ Iterations of the Darboux transformation starting from
the potential $u_0 =0$ and $f(x) =x$ give all exactly
solvable one-dimensional rational potentials \cite{AMM}. Note that
all these potentials are singular.

{\sc General construction scheme.} Take two Moutard transformations
of the Schr\"odinger operator with some potential $u_0$ defined by
the solutions $\omega_1$ and $\omega_2$ and denote the resulting
potentials $u_1$ and $u_2$ respectively. The transformed functions
$M_{\omega_1}(\omega_2)$ are defined up to terms $C/\omega_1$ and
satisfy the equation $(-\Delta+u_1)\psi=0$. The same holds for the
transformed  functions $M_{\omega_2}(\omega_1)$. Fix a function
$\theta_1$ from the set $M_{\omega_1}(\omega_2)$  and take the
function $\theta_2 = - \frac{\omega_1}{\omega_2}\theta_1$ from the
set $M_{\omega_2}(\omega_1)$. Now apply to the operator
$(-\Delta+u_1)$ the Moutard transformation defined by the function
$\theta_1$ and similarly the Moutard transformation defined by $\theta_2$ to the
operator $(-\Delta + u_2)$. The following result is well known:

\begin{lemma} $M_{\theta_1}(u_1) = M_{\theta_2}(u_2) = u$.
The functions $\psi_1 = \frac{1}{\theta_1}$ and $\psi_2 =
\frac{1}{\theta_2}$ satisfy the equation $L\psi = 0$, where $L =
-\Delta+u$.
\end{lemma}

In this construction we have a free scalar parameter for the choice
of the function $\theta_1$ from the set $M_{\omega_1}(\omega_2)$.
This parameter can be used in some cases to build a nonsingular
potential $u$ and the functions $\psi_1$ and $\psi_2$.

{\sc Example 1.} Let $u_0=0$, $\omega_1 = x +2(x^2-y^2)+xy$ and
$\omega_2 = x+y+\frac{3}{2}(x^2-y^2)+5xy$. Choosing $\theta_1$
appropriately (we omit its lengthy formula) one gets
\begin{equation}
\label{1} u = -\frac{5120    (1 + 8    x + 2y + 17 x^2 + 17
y^2)}{(160 + 4    x^2 + 4y^2 + 16    x^3  + 4    x^2y + 16    x
y^2  + 4    y^3 + 17(x^2+y^2)^2)^2},
\end{equation}
\begin{equation}
\label{2}
\begin{split}
\psi_1 = \frac{x + 2    x^2 + x    y - 2    y^2}{160 + 4    x^2 +
4y^2 +  16    x^3  + 4    x^2    y + 16    x    y^2 + 4
y^3 + 17 (x^2+y^2)^2}, \\
\psi_2 = \frac{2    x + 2y + 3    x^2 + 10    x    y - 3    y^2}{160
+ 4    x^2 + 4y^2 +  16    x^3  + 4    x^2    y + 16    x    y^2 + 4
y^3 + 17 (x^2+y^2)^2}.
\end{split}
\end{equation}

\begin{theorem}
The potential $u$ given by (\ref{1}) is smooth, rational  and
decays as $1/r^6$ for $r \to \infty$.

Functions $\psi_1$ and $\psi_2$ given by (\ref{2}) are smooth,
rational, decay as $1/r^2$ for $r \to \infty$ and span a
two-dimensional space in the kernel of the operator $L = -\Delta
+u: L_2(\R^2) \to L_2(\R^2)$.
\end{theorem}

{\sc Example 2.} Let $u_0=0$, $\omega_1 = x +
\frac{x^2-y^2-3xy}{5}+2(-x^3-3x^2y+3xy^2+y^3)$ and $\omega_2 = x+y
+\frac{x^2-y^2}{2}-\frac{xy}{5}-4(3x^2y-y^3)$. For some
appropriate $\theta_1$ one obtains a smooth rational potential $u$
decaying as $1/r^8$, as well as smooth rational $\psi_1$ and
$\psi_2$ in the $L_2$--kernel of the Schr\"odinger operator with
the potential $u$, $\psi_1$ and $\psi_2$ decay as $1/r^3$. We omit
here the explicit formulas.

We conjecture that increasing the degree of the initial harmonic
polynomials $\omega_1$ and $\omega_2$ one can for every $N>0$
construct potentials $u$ and their eigenfunctions $\psi_1$ and
$\psi_2$ decaying faster than $1/r^N$.

The authors would like to thank P.G.Grinevich and S.P.Novikov for
useful discussions.

\end{document}